\documentstyle[twocolumn,prl,aps]{revtex} 
\begin{document}
\draft
%
%%%%%%%%%%%%%%%%%%%%%%%%%%%%%%%%% TITLE PAGE
%

\title{Microscopic derivation of Kardar-Parisi-Zhang Equation}
\author{L. Bertini\cite{byline}}
\address{Imperial College, London U.K.}
\date{\today}
\maketitle
%
%%%%%%%%%%%%%%%%%%%%%%%%%%%%%%%%% ABSTRACT

\begin{abstract}
We consider the scaling limits for a one-dimensional random growth
model, the weakly asymmetric single step Solid-on-Solid process.
We show that the fluctuation field, if considered in an appropriate (long)
space-time scale, solves the Kardar-Parisi-Zhang equation. 
\end{abstract}
%
%%%%%%%%%%%%%%%%%%%%%%%%%%%%%%%%% PACS NUMBERS
%
\pacs{05.40.+j, 02.50.Ga, 68.10.Jy}
%
%%%%%%%%%%%%%%%%%%%%%%%%%%%%%%%%% PAPER CONTENT
%
\narrowtext

The problem of determining the macroscopic shape of growing processes
is a central topic in non-equilibrium Statistical Mechanics.
Most theoretical progresses have been made by analyzing oversimplified
models of the growth mechanism, which nonetheless catch some of its
essential physical aspects. A prototypical example of this approach is the 
so-called Eden model \cite{eden}.
A deeper understanding of the growth process involves the analysis of
its deviations from the most probable shape. These fluctuations
phenomena  take place on a finer scale, which we call here
mesoscopic. Accordingly, the
randomness in the microscopic dynamics is not completely averaged out by
the coarse-graining procedure, and manifests itself as a random term
in the mesoscopic equation describing the evolution of the fluctuations. 

The long scale behaviour (in mesoscopic units) of the 
fluctuations is believed to be {\sl universal}, i.e. independent, 
within a suitable class, on the particular growth model. 
The celebrated Kardar-Parisi-Zhang (KPZ) equation \cite{kn:kpz} has
been proposed exactly to describe this behaviour.
A striking feature of growth processes, which has been noted already by
Eden, is the roughness of the cluster surface; the so-called 
{\sl kinetic roughening}.
At the mesoscopic scale this is reflected by the presence, 
as compared to equilibrium phenomena, of large and non-Gaussian
fluctuations.
The KPZ model includes this effect by introducing 
a non-linear term in the dynamical equation. 

Having nothing to add on the much debated questions of the critical
indices and the upper critical dimension in the KPZ model, we address
here a different, more basic, problem.
We show how the KPZ equation, which is a mesoscopic (continuous)
model, can be obtained by considering the scaling limit for the
fluctuation field of a truly microscopic (discrete) random growth
model. The presence of non-Gaussian fluctuations, as it is clear in
the theory of critical phenomena, is a highly non trivial point. 
In this contest, the discrete model exhibits indeed Gaussian fluctuations
at the hydrodynamical scale, whereas they became non-Gaussian 
on an appropriate, longer space-time scale (in microscopic units). 
Although the discrete model is very peculiar this result confirms the
{\sl universality} assumption in the KPZ model.
In the analysis of the microscopic model we find 
that the limiting field solves the KPZ equation with a 
{\sl Wick renormalization} of the non linearity \cite{gianni}. 
Our discussion will also clarify the {\sl small gradient} hypotheses
which is behind any derivation of KPZ equation. 
We stress these results are obtained only in the one-dimensional
setting, i.e. for interfaces embedded in ${\bf R}^2$.  
The discussion is here kept at the physical level of rigour, 
the more mathematical inclined reader is referred to \cite{kn:bg} for
detailed proofs.

The microscopic model we analyze is the so-called {\sl Single Step
Solid-On-Solid} process \cite{sos}. It is an effective model 
for the separation line (interface) between two phases.
It represents the interface, in a local coordinate system, 
as a single valued function 
$\zeta: {\bf Z} \mapsto {\bf Z}$, where ${\bf Z}$ is the
one-dimensional lattice and $\zeta(x)$ is the interface
height at the point $x\in {\bf Z}$. The single step requirement is
imposed by restricting the allowed configurations to those
satisfying the constraint $|\zeta(x+1)-\zeta(x)|=1$. 
We introduce a {\sl weakly asymmetric} local random dynamics 
by the following growth rules: 
local minima become maxima, $\zeta_t(x)\mapsto \zeta_t(x)+2$,  with
rate $1/2+\varepsilon$ whereas 
local maxima become minima, $\zeta_t(x)\mapsto \zeta_t(x)-2$, 
with rate $1/2$. This occurs independently 
at each site and no other transition is allowed. The single step
constraint is then preserved by the time evolution. 
This process describes a local evaporation/deposition 
and introduces, for $\varepsilon>0$ a growth direction.
We also note this process can be obtained from the Metropolis dynamics
for a two-dimensional Ising model in the limit where the inverse
temperature $\beta\rightarrow\infty$ and the (positive) external
magnetic field $H\rightarrow 0$ keeping constant the product $\beta H
\propto \varepsilon$, see \cite{kn:ks}. In this limit the overhangs
are not allowed and the interface layer is sharp. In order to obtain 
the single step dynamics also the initial configuration has to be chosen
accordingly, e.g. one can take the {\sl plus} phase below the diagonal
and the {\sl minus} phase above.

The above dynamics can be represented in terms
of the {\sl Weakly Asymmetric Exclusion Process} (WASEP). Let us
consider a lattice gas with hard core interaction (at most one
particle per site) and let the particles jump to the left
(resp. right) with rate $1/2+\varepsilon$ (resp. $1/2$). We denote by 
$\eta = \{ \eta(x)\,,\, x\in {\bf Z} \}$ the particles configuration, 
$\eta(x)\in \{ 0,1 \}$. To WASEP is naturally associated a Markov
semigroup $e^{t L_\varepsilon}$ which acts on the functions on the
configuration space $\Omega= \{0,1\}^{\bf Z}$. The kernel of the
semigroup, $e^{t L_\varepsilon}(\eta,\eta')$, gives then the
transition probability  
for the transition $\eta\mapsto \eta'$. The process is however best
characterized by its {\sl infinitesimal generator} $L_\varepsilon$,
given by
\begin{equation}\label{le}
L_\varepsilon = L_0 + \varepsilon \: L^-
\end{equation}
where $L_0 = (L^+ + L^-)/2$ is the symmetric part and
\begin{equation}\label{l+-}
L^\pm f (\eta)  = \sum_x \eta(x) [ 1 - \eta(x\pm 1)] 
\: [ f(\eta^{x,x\pm 1}) - f(\eta) ]
\end{equation}
are the generators of the totally asymmetric dynamics. In the above
equation $\eta^{x,y}$ represents the configuration obtained from
$\eta$ by letting a particle jump from $x$ to $y$ and $f$ is a local
function on $\Omega$. Eq. (\ref{l+-}) is simply a formal way of
writing that in a small time interval $dt$ a particle jumps from $x$ to
$x\pm 1$ with probability $dt$ {\sl provided} there is a particle in
$x$ and the site $x\pm 1$ is empty. 
It is now a simple check to verify that the interface dynamics above
described can be obtained from WASEP by setting \cite{tag}
\begin{equation}\label{map}
\zeta_t(x) \equiv \sum_{y \leq x} \: [ 2\eta_t(y) -1 ]
\end{equation}

We now introduce a macroscopic coordinate $r=\varepsilon x$ such that
the strength of 
asymmetry coincides with the scale parameter. In this coordinate
system the interface position is given by
\begin{equation}
m^\varepsilon_\tau (r) = \varepsilon ~ \zeta_{\varepsilon^{-2}\tau} 
(\varepsilon^{-1} r) 
\end{equation}
where $\zeta (x)$ is defined by linear interpolation for non 
integer $x$. Above we scaled the
microscopic coordinate $x$, the interface height $\zeta$ and the
microscopic time $t$ diffusively, i.e. $x\sim \zeta\sim
\varepsilon^{-1}$, $t\sim \varepsilon^{-2}$. 
The results in \cite{wasep} imply then the {\sl hydrodynamical
limit} \cite{kn:s}. Assuming that $m^\varepsilon_0(r)$ converges, as
$\varepsilon\rightarrow 0$, to a continuous function $m_0(r)$, then 
$m^\varepsilon_\tau(r)\rightarrow m_\tau(r)$ which solves 
\begin{equation}\label{he}
\partial_\tau m_\tau = {1\over 2} \Delta m_\tau + 
{1\over 2} \left[ 1 - (\nabla m_\tau )^2 \right]
\end{equation}
where $\Delta$ (resp. $\nabla$) is the Laplacian (resp. the gradient).
Eq. (\ref{he}) describes the interface displacement as given by
three factors. A diffusion term which takes into account the local 
relaxation, a constant growth, and a non-linear term reflecting the
single step constraint: no growth is allowed where 
$|\nabla m(r)|=1$. 
We also note that, since $|\nabla m_0|\leq 1$, Eq. (\ref{he})
implies $|\nabla m_\tau|\leq 1$ for all $\tau >0$, as it is 
obvious in the discrete model.

The hydrodynamical limit is to be interpreted as a {\sl Law of Large
Numbers\/}: the 
microscopic dynamics is random, but in the scaling limit
$\varepsilon\rightarrow 0$ the deterministic equation (\ref{he}) is
obtained. Associated with it there is  a {\sl Central Limit
Theorem} (CLT) which is stated in the following form. Introduce the
interface fluctuations as
\begin{equation}
u^\varepsilon_\tau (r) = \sqrt{\varepsilon} ~ \left[
\zeta_{\varepsilon^{-2}\tau} ( \varepsilon^{-1} r) 
- {\bf E} \: \zeta_{\varepsilon^{-2}\tau}
(\varepsilon^{-1} r) \right]
\end{equation}
where ${\bf E}$ denotes the expectation value and the normalization 
$\sqrt{\varepsilon}$ is the usual one in CLT. Under some assumption 
on the initial configuration (basically the convergence holds at 
time 0), $u^\varepsilon_\tau$ converges to the solution of the {\sl
linear} Stochastic Partial Differential Equation (SPDE)
\begin{equation}\label{gf}
\partial_\tau u_\tau = {1\over 2} \Delta u_\tau 
- \nabla m_\tau \,  \nabla u_\tau
+ \sqrt{1 - (\nabla m_\tau )^2} \: {\dot w}_\tau
\end{equation}
where ${\dot w}_\tau(r)$ is the space-time white noise, i.e.
\begin{equation}
{\bf E} \left\{ {\dot w}_\tau (r) \, {\dot w}_{\tau'} (r') 
\right\} = \delta(\tau -\tau') \, \delta(r -r') 
\end{equation}
We stress that in (\ref{gf}) the function $m_\tau$ is given and it is
precisely the solution of (\ref{he}). The form of equation (\ref{gf})
is easily  motivated. The deterministic part is in fact the 
linearization of
the hydrodynamical equation (\ref{he}) around the trajectory
$m_\tau(r)$ whereas the prefactor of the noise is
again due to the single step constraint.

The results in \cite{wasep} thus rule out the possibility of observing
non-Gaussian fluctuations, as predicted by the KPZ equation, at the
hydrodynamical scale. We argue that this will be instead the case when
fluctuations on longer scale are considered. This possibility has been
already successfully pursued for some spin-flip (Glauber),
one-dimensional, long-range (mean-field type) models at the critical
temperature \cite{nem}. 

In the discussion on the hydrodynamical limit and the Gaussian
fluctuations, the condition on the initial configuration were rather
general; for instance one can choose a random $\eta_0$ distributed with
a slowly-varying product measure such that
\begin{equation}
{\bf E} \left\{ \eta_0 (x) \right\} = \rho_0 (\varepsilon x)
\end{equation}
and then take $\zeta_0 (x)$ as given by (\ref{map}). In this case one
obtains $m_0$ such that $\nabla m_0 = 2 \rho_0 -1$ and a Gaussian
process for $u_0$; its variance is determined by $\rho_0$.
In order to derive the KPZ equation we have instead to prepare
more carefully the initial state \cite{emy}. 
We assume $\zeta_0(x)$ to be a small perturbation of a 
straight line, furthermore the perturbation is so slowly-varying that
it changes only on the scale $\varepsilon^{-2}$, which is much longer
than the hydrodynamical scale previously discussed. 
Denoting by ${\overline \zeta}_0(x) = x \, \tan \theta$ the straight
interface we thus assume there exists an H\"older continuous function 
$h_0:{\bf R}\mapsto {\bf R}$ (it may be random) such that
\begin{equation}\label{is}
\varepsilon ~ \left[ \zeta_0(x) - {\overline \zeta}_0 (x) \right]
\approx  h_0 (\varepsilon^2 x)
\end{equation}
where the approximate identity means that for any $r\in {\bf R}$  
\begin{equation}\label{is2}
\lim_{\varepsilon \rightarrow 0} \varepsilon 
\left\{ \zeta_0( \varepsilon^{-2} r) - 
{\overline \zeta}_0(\varepsilon^{-2} r)
\right\} = h_0 (r)
\end{equation}
This is the precise formulation, in our contest, of the {\sl small
gradient} condition in the KPZ model.

A naive convergence result will then be the following. 
Assuming (\ref{is}), the interface position at time $t$ is 
\begin{equation}\label{dkpz}
\zeta_t (x + \varepsilon t \tan \theta ) \approx {\overline \zeta}_t 
(x+ \varepsilon t \tan \theta  ) 
+ \varepsilon^{-1} h_{\varepsilon^4 t} (\varepsilon^2 x) 
\end{equation}
where ${\overline \zeta}_t$ is the evolution of the straight interface, i.e. 
${\overline \zeta}_t(x) = {\overline \zeta}_0 (x) 
+ \varepsilon \, t \, (1-\tan^2 \theta) /2$, and $h_\tau=h_\tau(r)$ is
the solution of the KPZ equation
\begin{equation}\label{kpz}
\partial_\tau h_\tau = {1\over 2} \Delta h_\tau - 
{1\over 2} ( \nabla h_\tau)^2 + \sqrt{1-\tan^2\theta} \: {\dot w}_\tau
\end{equation}
We stress the fluctuation normalization in (\ref{dkpz}), even if the
non-linear Eq. (\ref{kpz}) is obtained, is still the CLT 
({\sl non anomalous}) normalization. We also note the space-time
scaling is  still diffusive, $x\sim\varepsilon^{-2},
t\sim\varepsilon^{-4}$. As usual \cite{kn:kpz} a Galilean
transformation in (\ref{dkpz}) is performed to obtain the KPZ equation.

The argument leading to the KPZ equation is the following.  
By Eq.s (\ref{le}), (\ref{l+-}) and (\ref{map}), we have
\begin{equation}\label{lez}
L_\varepsilon \zeta (x) = {1\over 2} (1+\varepsilon) \nabla^+ \nabla^-
\zeta(x) 
+  {\varepsilon\over 2} \left[ 1 - \nabla^+ \zeta(x) \nabla^- \zeta(x)
\right] 
\end{equation}
where $\nabla^\pm \zeta (x) = \pm [ \zeta (x\pm 1)- \zeta (x)]$ are
the discrete gradients and $\nabla^+\nabla^-$ the discrete Laplacian.
Inserting the {\sl ansatz} (\ref{dkpz}) in (\ref{lez}) and dropping 
higher order in $\varepsilon$, one obtains the 
deterministic part on the right hand side of (\ref{kpz}). 
Analogously, the noise term can be identified by the computation
\begin{eqnarray*}
\lefteqn{
L_\varepsilon \left[ \zeta (x) \zeta (y) \right] - \zeta(x)
L_\varepsilon \zeta(y) -
\zeta (y) L_\varepsilon \zeta (x)\phantom{merda}}\\
& = & \delta(x-y) 
\left\{  1 -\nabla^+ \zeta(x) \nabla^- \zeta(x) \right. \\
&&\left. 
+ \varepsilon \left[ 1 + \nabla^+ \nabla^- \zeta(x) - 
\nabla^+ \zeta(x) \nabla^- \zeta(x \right] \right\}
\end{eqnarray*} 

The main point in the above formal argument is of course the
validity of (\ref{dkpz}) under the assumption (\ref{is}), which is a
{\sl propagation of chaos} result. For WASEP this has been proven in a
strong form \cite{wasep}, but those results do not hold in the time scale we
are interested in. In fact we implicitly assumed $h_\tau (r)$ to be a
smooth (differentiable) function, but a simple analysis of
(\ref{kpz}) \cite{nd}, shows this cannot be the case and that
$r\mapsto h_\tau(r)$ can be at most $\alpha$-H\"olderian, $\alpha<1/2$. 

As emphasized in \cite{kn:bc}, a correct way to define the process
$h_\tau$ is through the Cole-Hopf transformation. 
For notation simplicity we discuss next only the case of a flat
interface, $\theta=0$, so
that no Galilean transformation in (\ref{dkpz}) is needed.
Let $\psi_\tau$ be the solution of the {\sl stochastic heat equation}
\begin{equation}\label{she}
\partial_\tau \psi_\tau = {1\over 2} \Delta \psi_\tau +  \psi_\tau 
\, {\dot w}_\tau
\end{equation}
which is linear and (in one space dimension) completely meaningful in
the space of continuous functions when the stochastic
differential is interpreted in the Ito sense. Furthermore, if the
initial condition is positive, then the process is strictly positive
for any $\tau>0$ as proven in \cite{mu}.
We can then {\sl define} 
(rigorously) $h_\tau = \log \psi_\tau$, which (formally) solves the
KPZ equation with a Wick renormalization of the non linearity, i.e.
\begin{equation}\label{kpzr}
\partial_\tau h_\tau = {1\over 2} \Delta h_\tau - 
{1\over 2} :\! ( \nabla h_\tau)^2 \!: 
+ {\dot w}_\tau
\end{equation}
where the Wick product can be informally written  as
$:\! ( \nabla h_\tau)^2 \!: \: = \: ( \nabla h_\tau)^2 - \delta(0)$.
The extra term $\delta(0)$ arises because Ito's rules of stochastic
calculus are to be used.

It is a fortunate coincidence that an analogous of the Cole-Hopf
transformation, first introduced by G\"artner \cite{wasep}, 
exists also for WASEP. 
Given $\zeta_t$ let us define a transformed process $\xi_t$ by 
\begin{equation}\label{dxi}
\xi_t(x) = \exp\{- \gamma_\varepsilon \zeta_t(x) 
+ \lambda_\varepsilon t \}
\end{equation}
where $\gamma_\varepsilon =  \log \sqrt{ 1 + 2 \varepsilon}$ and 
$\lambda_\varepsilon = 1 + \varepsilon - \sqrt{1+ 2 \varepsilon}$.
By a straightforward computation, it is verified that $\xi_t$
solves the stochastic equation 
\begin{equation}\label{semimart}
d \xi_t(x)  =  {e^{\gamma_\varepsilon} \over 2}  \: \nabla^+ \nabla^- 
\xi_t(x) dt + dM_t(x)
\end{equation}
where the first (linear) term is just the discrete Laplacian
and the noise term $M_t$ is characterized by
\begin{eqnarray}\label{ang}
\lefteqn{
{d\over dt} \:   \langle M (x) , M (y) \rangle_t
= 2 \, \varepsilon^2  \, \delta(x-y) \: \xi_s (x)^2 }\\
&\times& 
\left[ \eta_s(x) (1-\eta_s(x+1)) + e^{-2\gamma_\varepsilon} \eta_s(x+1) 
(1-\eta_s(x))\right]
\nonumber
\end{eqnarray}

The problem of showing the convergence of the interface fluctuations to
the KPZ equation is thus reduced in proving the convergence of 
$\xi_{\varepsilon^{-4}\tau} (\varepsilon^{-2}r)$ to $\psi_\tau(r)$. 
Since the first, linear, term is precisely the discretization of the
Laplacian in (\ref{she}), it comes for free. 
To identify also the noise, one has to show, using the basic hypotheses
(\ref{is}), that the term in the square bracket on the right
hand side of (\ref{ang}) converges to $1/2$. As $\eta(x)$ is a small 
perturbation of the product measure with density $1/2$ this is
intuitively clear and it can be rigorously justified \cite{kn:bg}.

By taking the inverse Cole-hopf transform we get
\begin{equation}\label{dkpzv}
\zeta_t (x) \approx  V_\varepsilon t + \varepsilon^{-1}
h_{\varepsilon^4 t} (\varepsilon^2 x)
\end{equation}
where 
$V_\varepsilon = \lambda_\varepsilon  / \gamma_\varepsilon 
= \varepsilon / 2 - \varepsilon^3 / 4! + O(\varepsilon^4) 
$
and $h_\tau(r)$ is now associated with the renormalized KPZ equation 
(\ref{kpzr}). 

A few comments are due. We first address the Wick
renormalization. We note that the product measure with a constant
density $1/2$ is stationary for WASEP, by (\ref{map}) 
$\zeta_t$ has then a steady state. The expectation of the non
linear part in (\ref{lez}) in this steady state is 
\begin{eqnarray*}
\lefteqn{
{\bf E}_{1/2} \left\{ \nabla^+ \zeta_t(x) \nabla^- \zeta_t(x) \right\}
}\\
& = & {\bf E}_{1/2} \left\{ ( 2 \eta_t (x) -1 ) ( 2 \eta_t (x+1) -1 )
\right\} =0 
\end{eqnarray*}
which is the same property that defines the Wick product. 
This shows very clearly that the non linearity in the KPZ equation 
cannot really be simply $(\nabla h_\tau )^2$, because the latter 
would have a positive expectation.
 
In a more general framework 
the transition from a microscopic description to the macroscopic
equation is carried out by using some  {\sl local
equilibrium} hypotheses (proven to hold in number of different models)
which allows to replace microscopic averages 
(in this case $\varepsilon \: \zeta_t(x)$) by macroscopic quantities 
($h_\tau(r)$). We have here discussed a situation where a naive local
equilibrium fails, but it is sufficient a minimal modification of 
the macroscopic term, i.e. the Wick product, to get the
correct answer. In our approach this fact is somehow hidden in the
Cole-Hopf transformation, a direct argument would be much neater and
helpful in suggesting how to deal with the general cases where no
Cole-Hopf transformation is available.

Comparing (\ref{dkpzv}) with (\ref{dkpz}) we see the extra term 
$\varepsilon^3/4!$ (which is finite in the scaling limit) in the drift
velocity $V_\varepsilon$. To explain it we have to
recall the well-known connection between the KPZ equation and the directed
polymers in a random environment. The solution of the stochastic heat
equation (\ref{she}) can in fact be interpreted as the partition
function of directed polymers in the quenched random environment
described by the potential $\dot w$. Accordingly, the quenched free
energy per unit of length is given by \cite{kar}
\begin{equation}
\lim_{\tau\rightarrow\infty} { {\bf E} \left\{ - \log \psi_\tau \right\}
\over \tau } =  {1\over 4!}
\end{equation}
We now recall that the solution of KPZ equation has been defined by
$h_\tau= \log \psi_\tau$. Since the expectation of $\zeta_t$ in the 
steady state is given by 
${\bf E}_{1/2} \{ \zeta_t (x)\} = \varepsilon \, t /2$, 
we see that the second term in $V_\varepsilon$ is indeed needed to get
the correct result.
We can also view the above discussion as a non-replica argument (but
still not a rigorous proof) to get the exact value of the 
quenched free energy.

We finally comment on the most unsatisfactory point in our approach, that
is the fact $\varepsilon$ is {\sl both} the scale parameter {\sl and} the
asymmetry strength. It would be tempting to try the asymmetric exclusion
process, where the asymmetry is kept fixed and independent on
$\varepsilon$.  
However in this case the correction to the hydrodynamic (which is
here the inviscid Burgers equation in the Euler scaling) 
are expected to appear at time $\varepsilon^{-3/2}$, 
before the random noise in the dynamics shows up \cite{kn:s}. 
In two dimension logarithmic corrections are
instead expected. In three or more space dimension there is a rigorous 
analysis beyond the hydrodynamical scale \cite{emy} and 
the limiting equation is deterministic. 
The interpretation in term of the interface dynamics is anyhow purely
one-dimensional.

To summarize, we have analyzed a one-dimensional growth process beyond
its first nontrivial hydrodynamical scale (which is here the diffusive
scale) and found  a scaling limit where the fluctuations evolve exactly 
according to the KPZ equation. In the Renormalization Group terminology
this can be stated saying that the {\sl irrelevant} term in the
microscopic dynamics do not survive the limit, thus proving the {\sl
universality} assumption.
In a more precise formulation, which takes full advantage of the
Cole-Hopf transformation, the Wick renormalization of the non-linearity
appears naturally.

%\acknowledgments

It is a great pleasure to thank N. Cancrini and G. Gia\-co\-min for the
long lasting collaboration. Most useful conversations with 
F. Dunlop, G. Jona-Lasinio, M. Marsili, E. Presutti and H.T. Yau
are also acknowledged. I am grateful to the 
Centre de Physique Theorique, 
Ecole Polytechnique, Palaiseau, where part of this work has
been carried out.

%%%%%%%%%%%%%%%%%%%%%%%%%%%%%%%%% REFERENCES LIST


\begin{references}
\bibitem{tag}
The infinite sum on $y$ is of course purely formal, it can be replaced 
by a finite sum introducing a {\sl tagged} particle from which start the
count.
\bibitem{nd}
It is enough to consider the associated linear equation.
\end{references}
\end{document}